\documentclass{mn2e}
\input epsf.sty

\def\tmin{t_{\rm min}}
\def\tmax{t_{\rm max}}
\def\rtr{r_{\rm tr}}
\def\Rtr{R_{\rm tr}}
\def\Rg{R_{\rm g}}
\def\Ns{N_{\rm s}}
\def\Or{\Omega/2\pi}

\def\lh{l_{\rm h}}
\def\lhmax{l_{\rm h, max}}
\def\lsoft{l_{\rm s}}
\def\vff{v_{\rm ff}}
\def\ttrav{t_{\rm trav}}
\def\Rms{R_{\rm ms}}
\def\Rout{R_{\rm out}}
\def\Rin{R_{\rm in}}
\def\rms{r_{\rm ms}}
\def\rout{r_{\rm out}}
\def\rin{r_{\rm in}}
\def\taues{\tau_{\rm es}}
\def\Rflare{D_{\rm flare}}
\def\Fcont{F_{\rm cont}}
\def\Fline{F_{\rm line}}
\def\sigmaT{\sigma_{\rm T}}
\def\me{m_{\rm e}}
\def\kT{k T_{\rm e}}
\def\MSun{{\rm M}_{\odot}}
\def\kapabs{\kappa_{\rm abs}}
\def\kapes{\kappa_{\rm es}}
\def\Sref{S_{\rm refl}}
\def\Sprim{S_{\rm prim}}
\def\Ka{K$\alpha$\ }

\title[Variability of the Fe K$\alpha$ line]
{Modelling the variability of the Fe K$\alpha$ line in accreting black holes}

\author[P. T. \.{Z}ycki]{Piotr T. \.{Z}ycki\thanks{e-mail: ptz@camk.edu.pl} \\
    Nicolaus Copernicus Astronomical Centre, Bartycka 18, 00-716 Warsaw,
Poland}

\date{18 March 2004}

\begin{document}
\label{firstpage}

\maketitle

\begin{abstract}
The variability of the Fe \Ka line near 6.5 keV seems to be reduced compared to 
the variability of the hard X-rays which presumably drive the line emission.
This is observed both in active galactic nuclei and galactic black hole binaries.
We point out that such reduced variability, as well as lack of coherence
between the variations of the line and the continuum, are a natural prediction
of a propagation model of variability in the geometry of inner hot accretion
flow. 
We compute detail model predictions of the variability characteristics which
could be compared with current and future data. We also point out that the model
requires a gradual disappearance of the cold disc, rather than a sharp transition
from the cold disc to a hot flow.

\end{abstract}


\section{Introduction}

The Fe \Ka fluorescent/recombination line near 6.5 keV is an important diagnostic
of accretion flows around compact objects (see Reynolds \& Nowak 2003 for a recent 
review). It is the strongest hard X-ray 
($E>1$ keV) line which can originate in the innermost regions of accretion flows
($\le 100\,\Rg$). It is indeed observed in energy spectra from all kinds of accreting 
sources: 
black hole and neutron star X-ray binaries, cataclysmic variables, active galactic
nuclei (AGN).

The line is produced when plasma is irradiated by hard X-ray ($E > 7$ keV)
radiation. If the plasma is  Thomson thick the continuum spectral component 
formed as a result of the irradiation (``Compton reflection'') has a 
characteristic shape, peaking
at 20--30 keV (Lightman \& White 1988). The line and the Fe K-shell absorption edge
are superposed on this continuum (George \& Fabian 1991; Matt, Perola \& Piro 1991). 
The properties
of the line and edge depend on the ionization of the reflecting medium:
with increasing ionization the line and edge shift towards higher energies,
while their strength increase  (e.g.\ \.{Z}ycki \& Czerny 1994 and references
therein). The profile of the line may be modified by relativistic and
kinematic effects, if the line originates in e.g.\ a rotating accretion disc
(Fabian et al.\ 1989). The line and absorption edge are then broadened
and smeared.
Such broad features are commonly seen in Seyfert galaxies (e.g.\ MCG-6-30-15, Tanaka
et al.\ 1995, Fabian et al.\ 2002)
and black hole binaries (BHB; e.g.\ Cyg X-1, Done \& \.{Z}ycki 1999;
GRS 1915+105, Martocchia et al.\ 2002),
providing a clear evidence for a relativistic accretion disk extending deep into
the gravitational potential of the central black hole.

X-ray emission from accreting sources is highly variable in a broad range
of time-scales. Most of the variability power is located in the range of Fourier
frequency $f \approx$(0.1--1)$M/(10\,\MSun)$ Hz, corresponding to a time-scale
($T=1/f$) of a few
seconds for a $10\,\MSun$ stellar black hole and a few weeks for a $10^7\,\MSun$
AGN (BHB review in McClintock \& Remillard 2003;
Markowitz et al.\ 2003b for AGN). Typical power density spectrum is roughly
a power law with slope $\alpha\approx 0$ ($P(f)\propto f^{\alpha}$) at low $f$, 
steepening to $\alpha\approx -1$ at $f \approx 0.1$ Hz and to $\alpha\approx -2$
at $f=1$--3 Hz, for the BHB in low/hard state. 
Typical root-mean-square (r.m.s.) variability is 20--30\%. 
The variability is stochastic rather than caused by deterministic chaos
type of process (Czerny \& Lehto 1997).

The observed variability of the Fe \Ka line and the entire reprocessed component
is somewhat surprising: they generally show rather less variability than the high
energy continuum which is presumably driving the line emission and Compton
reflection. This is seen both in AGN and BHB. In AGN the \Ka line seems either not
to respond to continuum variations on time-scales of minutes to days
(e.g.\ Reynolds 2000; Done, Madejski \& 
\.{Z}ycki 2000; Chiang et al.\ 2000), or the line variability
appears to be uncorrelated with that of the continuum (Vaughan \& Edelson 2001). 
In particular, studies of r.m.s.\ variability amplitude as
a function of energy demonstrate the reduced variability in a relatively model
independent way (Inoue \& Matsumoto 2001; Markowitz, Edelson \& Vaughan 2003a). 
The short term variability of the reprocessed component is more difficult to measure
because of insufficient statistics, but where such studies were possible, the results
also suggested reduced variability (e.g.\ Done et al.\ 2000).

A similar effect is seen in BHB systems. The time resolved spectral analysis
is more difficult for BHB than for AGN,  but Fourier spectroscopy (study of 
Fourier-frequency resolved spectra; Revnivtsev, Gilfanov \& Churazov 1999, 2001) 
can be used to investigate the
variability characteristics on short time scales. Indeed, the amplitude of 
the reprocessed
component in high Fourier frequency spectra is smaller compared
to that in low Fourier frequency spectra (Revnivtsev et al.\ 1999, 2001).
In other words, the reprocessed component responds to the
variability of the primary continuum on long time scales ($T=1/f \ge 1$ sec)
but it does not do so on short time scales.

In AGN one reason for the lack of variability might be the contribution to
the reprocessing by a distant matter, e.g.\ the obscuring dusty torus. Indeed,
a narrow Fe \Ka line is observed in many Seyfert galaxies.
However, it is the variability properties of the broad component (hence
produced very close to the central black hole) that are so puzzling. 

One suggested explanation invoked complex ionization effects in the illuminated
surface of the disc (e.g.\ Nayakshin, Kazanas \& Kallman 2000).
 Formation of a hot ionized skin (where the Fe \Ka line is not produced)
with thickness proportional to the X--ray 
flux may lead to anti-correlation between the line flux and the continuum
flux. This model was quantitatively tested by \.{Z}ycki \& R\'{o}\.{z}a\'{n}ska
(2001), who concluded that it does indeed predict certain decrease of amplitude
of variability of the line, although it is not possible to obtain an absolutely
constant line flux.

In this paper we demonstrate that the reduced variability of the Fe \Ka line
is a natural result from a propagation model of X-ray emission in the geometry
of a hot inner accretion flow. This geometry is one of the possibilities
for accretion flow in low/hard states of accreting black holes
(e.g.\ Di Salvo et al.\ 2001; see Done 2002 for review). 
It is supported by X--ray spectral studies  (review in Poutanen 1999), 
in particular the correlation between spectral slope and amplitude of
reflection (Zdziarski, Lubi\'{n}ski \& Smith 1999) and long time scale spectral
evolution of black hole binaries (Zdziarski et al.\ 2003). The physical mechanism
of formation the two-phase plasma flow may be related to plasma 
evaporation/condensation (R\'{o}\.{z}a\'{n}ska \& Czerny 2000 and references therein).
The idea of
propagating X-ray emitting structures was put forward by Miyamoto et al.\ (1988),
based on early variability studies of Cyg X-1. This was further developed and
tested by e.g.\ Nowak et al.\ (1999), Misra (2000), and formulated in a more
general form by Kotov, Churazov \& Gilfanov (2001). Recently, \.{Z}ycki (2003) showed that the 
Fourier-frequency resolved spectra can be reproduced in the propagation model,
while Uttley (2004) argued for this model based on the r.m.s.--flux relation in
accreting pulsar SAX J1808.4-3658 (see also Uttley \& M$^{\rm c}$Hardy 2001).

\section{The Model}
\label{sec:model}

The model was also described in detail in \.{Z}ycki (2003).
X--rays are assumed to be produced by
compact active regions/emitting structures traveling from outside inwards, 
towards the black hole.
The structures originate at a certain radius, $\Rout$, and they move towards
the centre at a fraction of the free-fall speed,
\begin{equation}
\label{equ:motion}
 v = \beta \vff = \beta \sqrt{ {2 G M \over R}},
\end{equation}
where $\beta<1$. The plasma heating rate is assumed to depend on radius
\begin{equation}
 \label{equ:lh}
 \lh(t) \propto R(t)^{-2}\,b[R(t)],
\end{equation}
where the exponent $-2$ corresponds to gravitational energy dissipation per ring 
of matter.
Here $l$ is the compactness parameter, 
$l\equiv (L/\Rflare) \sigmaT/(\me c^3)$
(where $\Rflare$ is the characteristic radius of the structure, assumed constant in time),
and $b(R)$ is the boundary term which is assumed to have its standard
form $b(R) = 1-\sqrt{6\Rg/R}$ (Shakura \& Sunyaev 1973). 

The heating rate
is normalized to have an assumed maximum value, $\lhmax$, the same for all flares. 
The equation of motion (Eq.~\ref{equ:motion}) can be solved to give
\begin{equation}
r(t) = (\rout^{3/2}-A t)^{2/3}, ~~~~~~~~~~~ 
                A \equiv {3\sqrt{2}\over 2} {\beta c \over \Rg}
\end{equation}
where the radial positions $r(t)$ and $\rout$ are expressed in units of $\Rg \equiv GM/c^2$
(lowercase $r$ will denote radial position in units of $\Rg$). 
The duration of a flare is
\begin{equation}
\label{equ:ttrav}
\ttrav=(\rout^{3/2}-\rin^{3/2})/A, 
\end{equation}
where the final radial position is  assumed $\Rin = 6\,\Rg$. In actual computations
we treat $\ttrav$ as a parameter, and solve for $\beta$. Following Poutanen \& Fabian
(1999) we generate $\ttrav$ according to a probability distribution 
$P(t) \propto t^p$ for $t$ between $\tmin$ and $\tmax$.


Soft photons for Comptonization are assumed to
come from reprocessing and thermalization of the hard X--rays, so that the feedback 
loop is realized as needed to explain the correlation between the amplitude
of reflection and spectral index (Zdziarski et al.\ 1999; Gilfanov, Churazov \& 
Revnivtsev 2000).  The geometrical scenario considered here is that of 
an inner hot flow partially overlapping a cold, optically thick disc disrupted at 
a certain  radius, $\Rtr$, (see e.g.\ Poutanen, Krolik \& Ryde, 1997, for arguments 
for the overlap). The luminosity of soft photons crossing an active region 
is parametrized as
\begin{equation}
 \label{equ:lsoft}
  \lsoft(t) = \Ns \lh(t) \times C[r(t)] = \Ns \lh(t) \times \left\{
 \begin{array}{cc}
             1                    & \mbox{for } r \ge \rtr \\
    \left({r \over \rtr} \right)^{\gamma}  & \mbox{for } r < \rtr,
 \end{array}
 \right.
\end{equation}
where $C(r)$ represents a covering
factor of the cold reprocessing matter. The covering factor is assumed to
increase with $R$ (i.e.\ $\gamma>0$), up to $C(\rtr)=1$. The normalization constant
$\Ns \approx 0.5$, as appropriate for a continuous corona at $r>\rtr$.
Thus, during a flare the Comptonized spectrum has a constant slope of $\Gamma\approx 2$
up to the moment of crossing the truncation radius, and then the spectrum get harder
($\Gamma$ decreases). The truncation radius and exponent $\gamma$ determine the average 
slope of the Comptonized spectrum. 

The emitting structures are assumed to originate at a mean
rate of $\lambda$ per second. Time intervals between their launch
are generated from the $\lambda \exp(-\lambda t)$ 
distribution, as appropriate for a Poissonian process. We adopt here
the flare avalanches description (Poutanen \& Fabian 1999), where
each spontaneous (``parent'') flare has certain probability to stimulate a 
number of ``baby'' flares. 
The stimulated flares are delayed after their parent flare (Poutanen \& Fabian 1999).

Each emitting structure is followed until it reaches the marginally
stable orbit at $\Rms = 6\,\Rg$. At each time step the primary Comptonized
component and its reprocessed component are computed. 

The  Comptonized spectrum is computed using the code
{\sc thComp} (Zdziarski, Johnson \& Magdziarz 1996), solving the Kompaneets
equation. Computations are parametrized by the photon spectral index,
which we compute from $\Gamma = 2.33(\lh/\lsoft)^{-1/10}$ (Beloborodov
1999a,b), and electron temperature 
$\kT/(\me c^2)$ computed using formulae from Beloborodov (1999b).
Plasma optical depth is assumed $\taues = 1$.

The spectrum of the Compton
reflected continuum is computed from the simple formula of
Lightman \& White (1988),
\begin{equation} 
 \label{equ:lwrefl} 
  \Sref(E) = {1-\epsilon \over 1+\epsilon} \Sprim(E), \quad\quad 
      \epsilon = \sqrt{ {\kapabs \over \kapabs + \kapes}}, 
\end{equation} 
where $\kapabs(E)$ and $\kapes(E)$ are the photo-absorption and electron 
scattering opacities, respectively, the former for a ``cold'' matter. 
The above formula is multiplied by a 
simple exponential cutoff to mimic the Klein-Nishina cutoff. 
The Fe \Ka line is added to the reflected continuum,
with equivalent width (EW) as a function of $\Gamma$, following computations
of \.{Z}ycki \& Czerny (1994). Specifically, we find
that the formula
\begin{equation}
EW(\Gamma) = 1.78\left(\Gamma/1.1\right)^{-0.5-0.15\,\Gamma}\ {\rm keV}
\end{equation}
reproduces the dependence of EW (relative to the reflected continuum)
on the spectral slope, $\Gamma$,
for cold matter with solar iron abundance, and disk inclination of $30^{\circ}$.

The instantaneous amplitude of the reprocessed component, $\Or$, should be related
to the covering factor of the cold plasma, $C(r)$ (Eq.~\ref{equ:lsoft}).
In practice we assume
\begin{equation}
 \label{equ:refl}
  \Or(r) = C(r) = \left\{
 \begin{array}{cc}
             1                         & \mbox{for } r \ge \rtr \\
        \left({r \over \rtr}\right)^{\gamma}& \mbox{for } r < \rtr,
 \end{array}
 \right.
\end{equation}
i.e.\ $\Or$ decreases from 1 at $r \ge \rtr$ to $\Or\ll 1$ at $r=\rms$. 

\begin{figure}
 \epsfxsize = 0.5\textwidth
 \epsfbox[18 230 620 710]{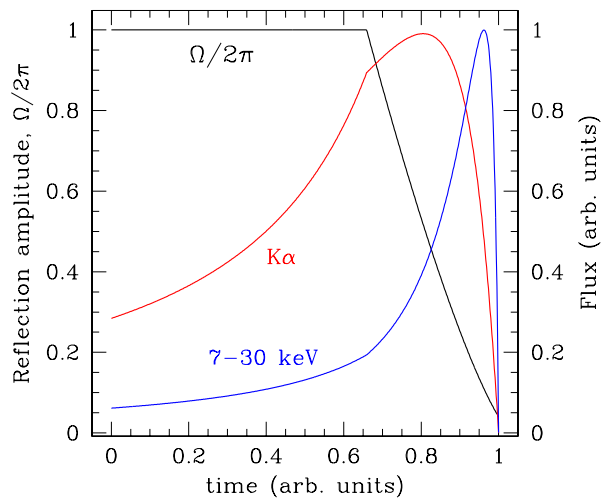}
 \caption{
During a flare the \Ka line flux  peaks {\em earlier\/} than the 7--30 keV 
continuum flux. Since the line flux can be written
$\Fline(t) \propto \Or(t)\times F(t)$, it decreases when the amplitude of 
reflection, $\Or$, decreases for $r<\rtr$. 
Note that the line and continuum fluxes were rescaled to 1 at the maximum, 
so the flare
amplitudes in these two energy bands are {\em not\/} represented accurately.
\label{fig:prof}}
\end{figure}

\begin{figure*}
 \epsfysize = 7 cm
 \epsfbox[18 350 620 710]{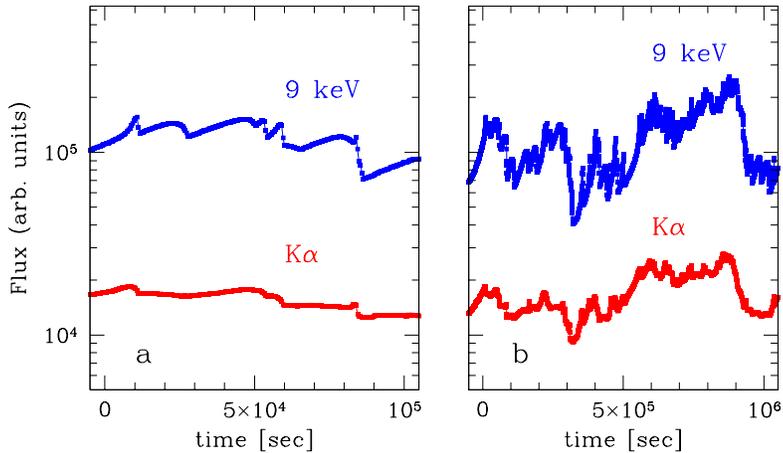}
 \caption{
Part of a typical light curve: 9 keV continuum and the \Ka line.
Panel (a): $10^5$ sec ($\approx 1$ day), panel (b): $10^6$ sec.
On time scale the variability of the line is clearly smoother (less 
high-frequency variations, smaller peak-to-peak amplitude) than of the 
continuum. This is still visible, although less so on the longer time scale
plotted. The black hole mass assumed for scaling the time scales is $10^7\,\MSun$.
\label{fig:lcurve}}
\end{figure*}

The sequence of spectra created by the above procedure is subject to standard 
analysis in the time and Fourier domains (see e.g.\ van der Klis 1995; 
Nowak et al.\ 1999; Poutanen 2001). 

All computations are performed assuming the central black hole mass $M=10^7\,\MSun$.
The time-scales of variability are assumed to follow a simple scaling with
the black hole mass. Consequently, the values of parameters of the variability model are adjusted
so that the power spectrum density (PDS) has the same shape as PDS of Cyg X-1 in low/hard 
state, and is shifted in frequency by the mass ratio. 
This is achieved for $\tmin = 5\times 10^{-3} M_1$ sec, $\tmax = 2 M_1$ sec, $p=-1$,
where $M_1 = M/(10\MSun)$.
The emitting regions are launched at $\Rout=100\,\Rg$ at a rate 
$\lambda = 40 M_1^{-1}\,{\rm sec^{-1}}$. The truncation radius is assumed $\Rtr=30\,\Rg$,
while $\gamma=2$.
We consider a stationary model, i.e.\ the above values are constant in time.

\section{Results}
 \label{sec:results}

\begin{figure*}
 \parbox{\textwidth}{
 \epsfxsize = 0.4\textwidth
 \epsfbox[18 160 620 710]{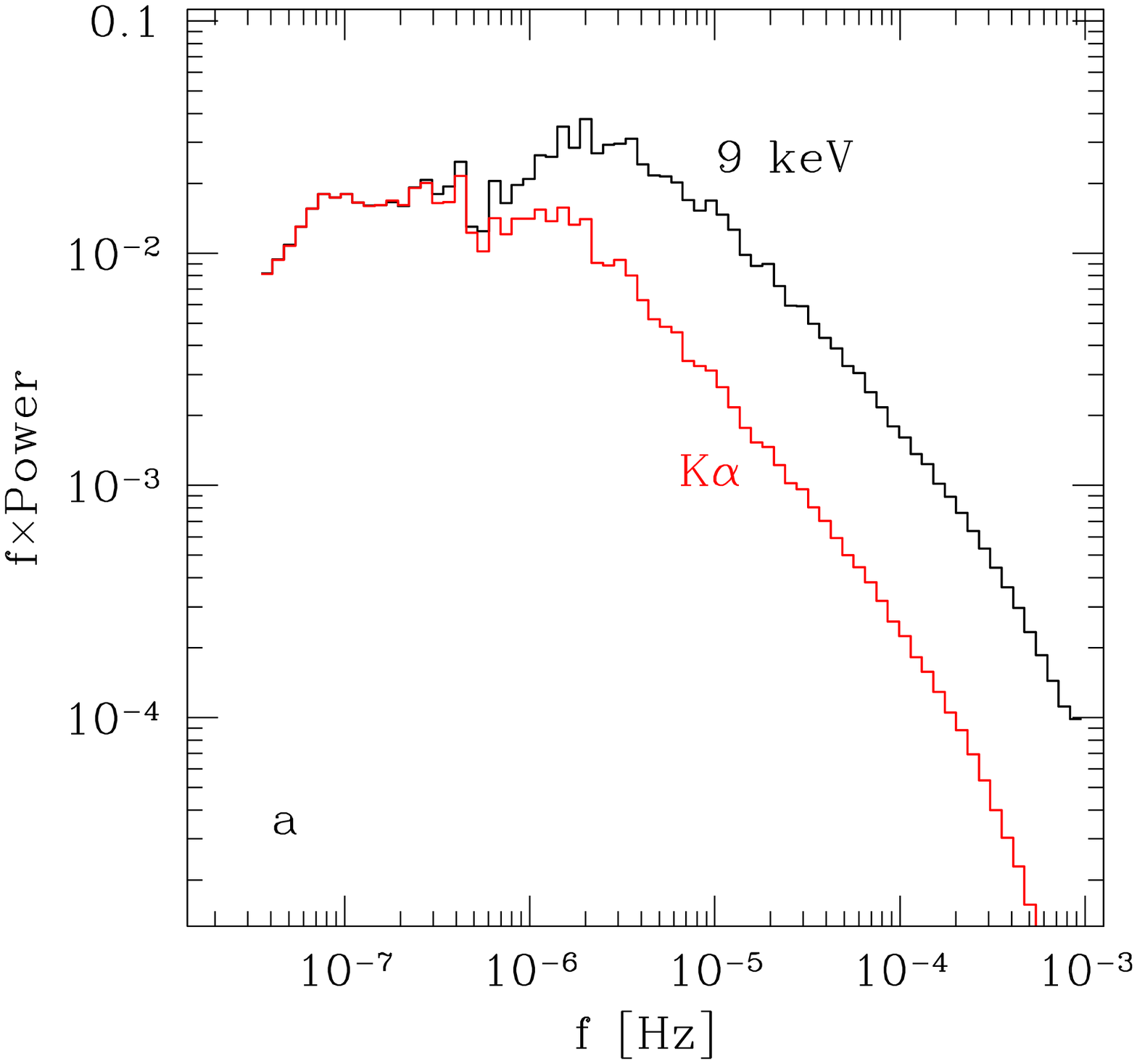}
 \epsfxsize = 0.4\textwidth
 \epsfbox[18 160 620 710]{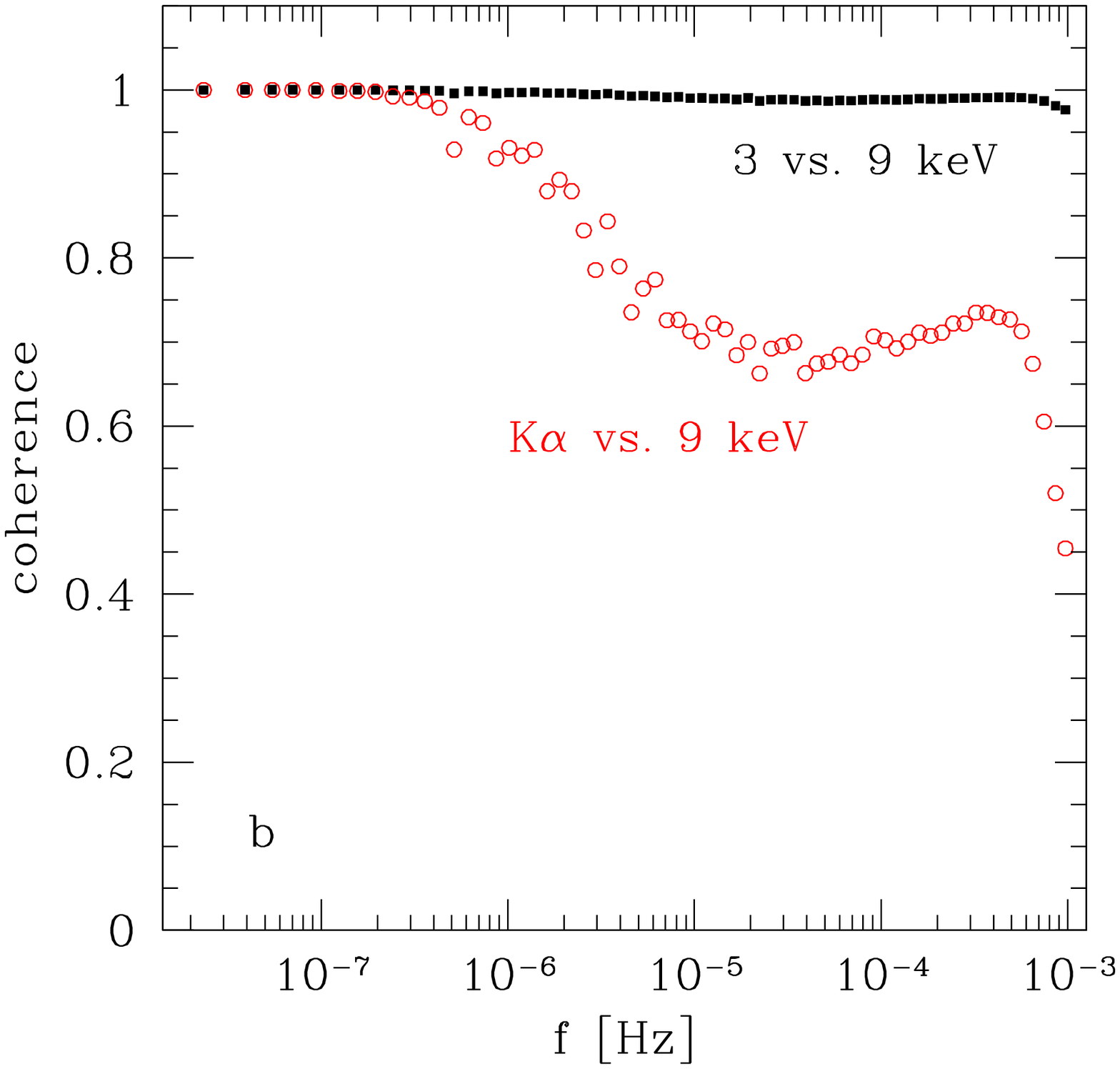}
}
 \parbox{\textwidth}{
 \epsfxsize = 0.4\textwidth
 \epsfbox[18 160 620 710]{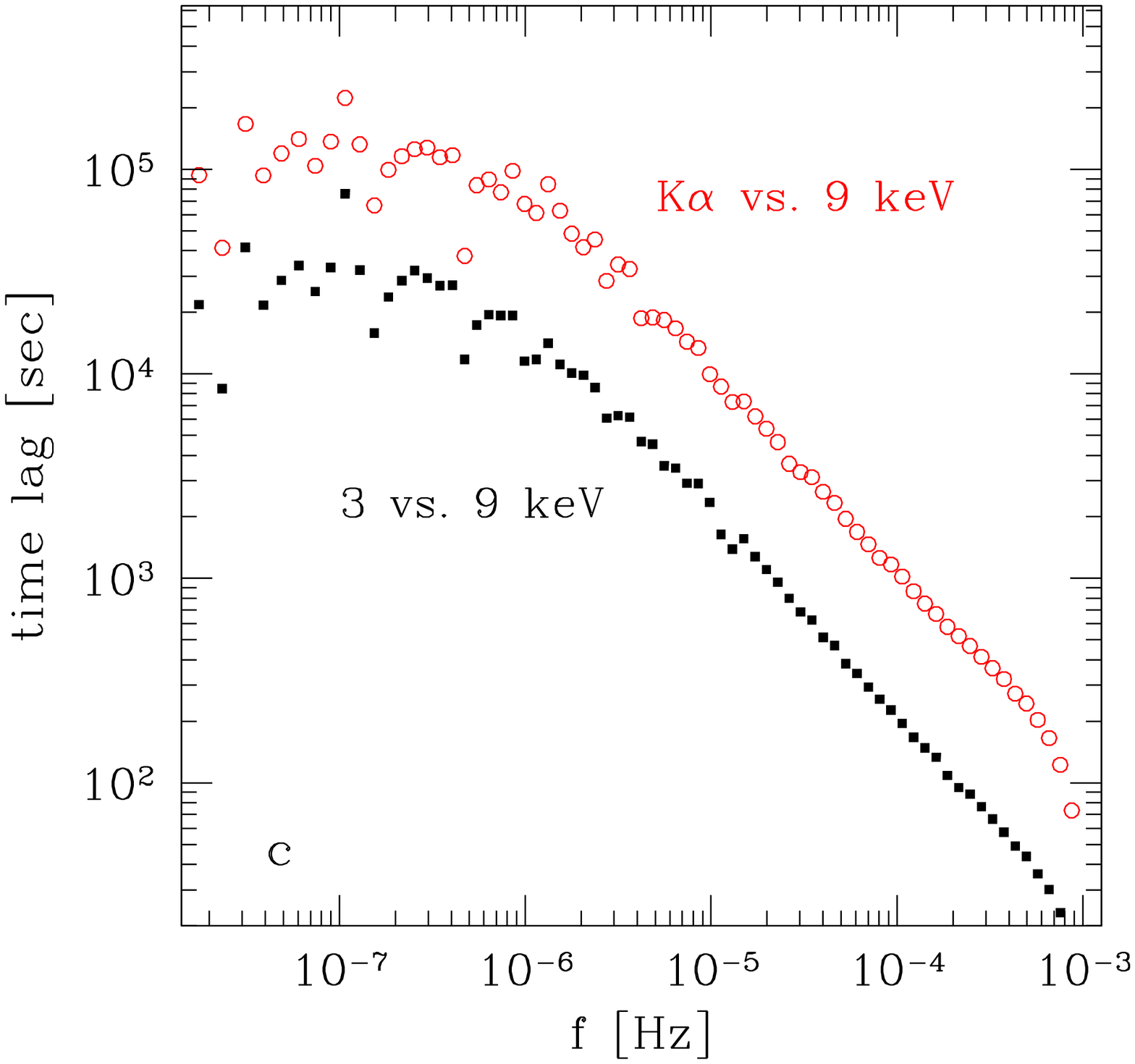}
 \epsfxsize = 0.4\textwidth
 \epsfbox[18 160 620 710]{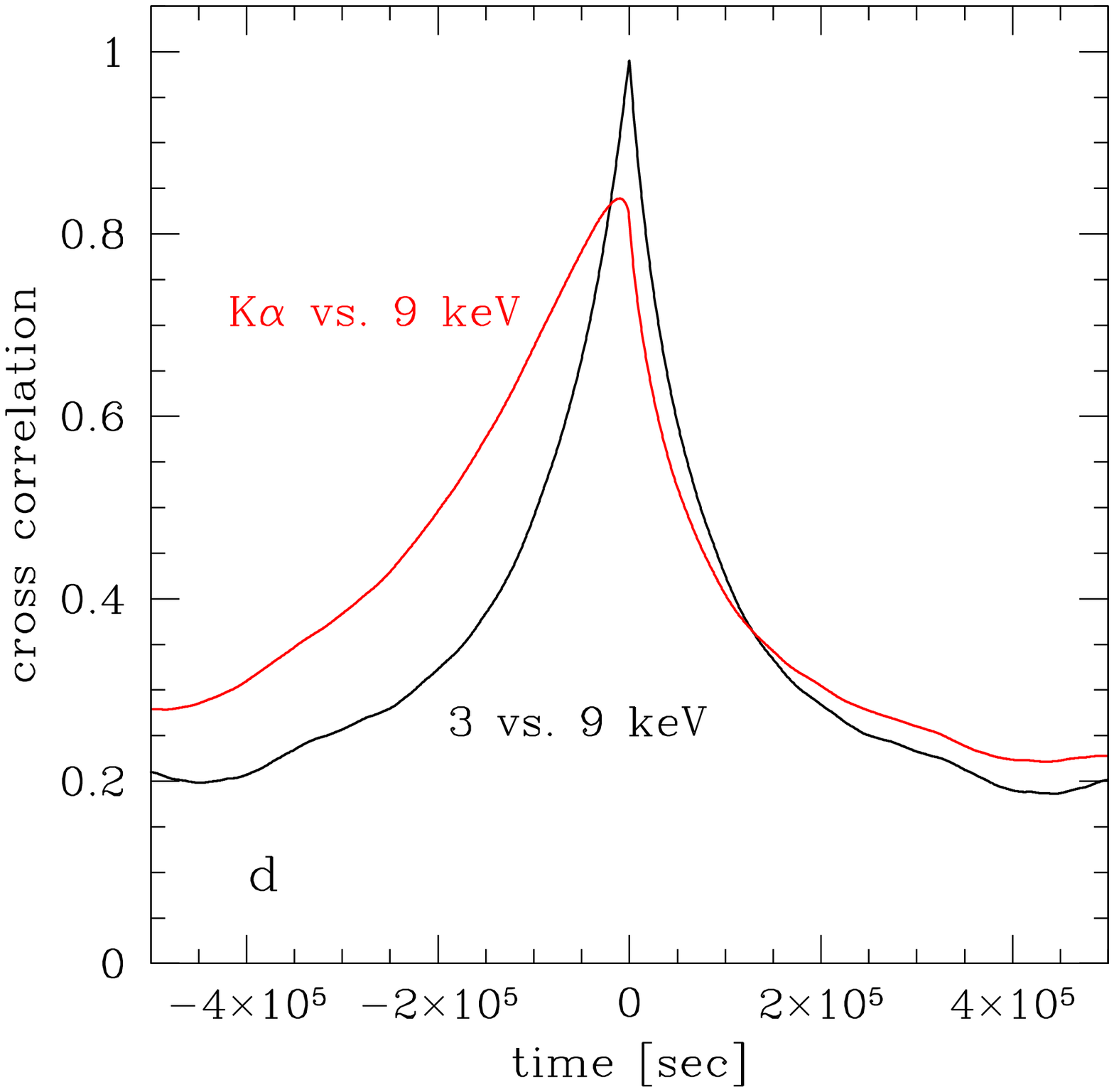}
}
 \caption{
Variability properties of the \Ka line vs.\ those of the 9 keV
continuum. (a) power spectra, (b) coherence function, (c) time lags
and (d) cross-correlation function. The \Ka line shows weaker high-frequency
variability compared to the continuum and there is a marked loss of coherence
between the line and the continuum, both effects occurring above the
frequency corresponding to longest flares. The time lags of the 9 keV continuum
relative to the line are much longer than those relative to the 3 keV
photons (factor of $\approx 5$), 
despite smaller energy separation. The lags can also be seen from
the highly asymmetrical cross-correlation function.
\label{fig:res4}}
\end{figure*}

The flux in the \Ka line (the line light curve) is found simply
by subtracting the continuum contribution from the total flux in the line bin.
The Fe \Ka line predicted by our model is narrow because of relatively
low velocities of the active regions. This is required to reproduce 
the relatively long time-scales of maximum power, but it is also a consequence of 
the assumption that the active regions move purely radially. 
If they possessed a significant azimuthal velocity component, their trajectories
would be longer spirals and the total velocity would have to be accordingly higher,
in order to produce the same flare durations. The azimuthal velocity component
would in that case give rise to Doppler broadening of spectral features, but
the variability properties of the model would remain unchanged.

Temporal profile of a typical flare is plotted in Fig.~\ref{fig:prof}. 
The profile of the heating rate, $\lh(t)$, and total Comptonized emission
are similar to the plotted profile of the 7--30 keV flux.
The amplitude of reflection, $\Or$, remains constant at 1 until the emission 
region
has reached the truncation radius, $\rtr$, and then $\Or$ begins to decrease.
This has an important consequence for the flux of the \Ka line: the line flux
peaks {\em earlier\/} than the continuum flux. 
This is simply because the line flux can be
expressed as the product of $\Or$ and $\Fcont$,
$\Fline \propto \Or\times \Fcont$. The magnitude of the difference in peaks 
position
depends on the exponent $\gamma$ describing the $\Or(r)$ dependence 
(Eq.~\ref{equ:refl}), and it is $\approx 0.2$ of the total flare duration for
the assumed $\gamma=2$. The second important consequence of the obtained $\Or(t)$
relation is that the amplitude of the flare in the \Ka line is smaller than
in the continuum (this is {\em not\/} shown in Fig.~\ref{fig:prof}).

Both the above effects have important consequences for line and continuum
light curves, 
examples of which are plotted in Fig.~\ref{fig:lcurve}. 
Variations of the line are clearly smoother that
those of the continuum: there is less high-frequency variability and
the peak-to-peak amplitude is clearly smaller in the line light curve,
although major events in the continuum are reflected in the line light curve.
The decoupling of variability is more clearly visible on the short term 
light curve ($10^5$ sec),
the differences between the line and continuum curves decreasing for
longer time-scales.
The curves are very similar when viewed on time scales of order
$\sim 10^7$ sec. Another way of demonstrating the reduction of variability
is to compute the power spectrum. This is plotted in 
Fig.~\ref{fig:res4}, together with a number of other characteristics.
The PDS of the line cuts off  above
$f_1 \approx 10^{-6}$ Hz much more rapidly than PDS of the continuum.
The r.m.s.\ computed for $f>f_1$ is $0.28$ for the continuum, while it is
only $0.16$ for the line. The frequency $f_1$ is related to the time-scale
of longest individual flares ($\tmax = 2\times 10^6$ sec in our computations). 

There is a clear lack of coherence between the line
and continuum light curves, as shown in panel (b) of Fig.~\ref{fig:res4}.
The coherence plotted there is defined as in Vaughan \& Nowak (1997),
and it is a measure of linear relation between two light curves. Generally, 
observed light curves  show coherence very close to 1
in a broad range of Fourier frequency (Nowak et al.\ 1999 for Cyg X-1; 
Vaughan, Fabian \& Nandra, 2003, for MCG-6-30-15). This indicates that each active 
region produces radiation
in the entire X--ray band, and so photons of different energies track each other's
variations very closely. Here, however, the line photons do not follow the
continuum photons in a simple way, because of the varying geometrical factor $C(r)$.
Hence the transformation between the continuum and line photons is
complicated and the coherence is reduced below unity.

The time delay between the line and continuum photons is much longer
than between the two continuum bands of similar energy separation
(the line is {\em leading\/} the variability of the continuum at $E>7$ keV).
Figure~\ref{fig:res4}c demonstrates this effect. Time lags between
the 6 keV and 9 keV continuum bands would be shorter than
the plotted time lag between 3 keV and 9 keV bands. However, the time lag
between the 6.4 keV \Ka line and the 9 keV band is actually longer
(by a factor of $\approx 5$) than the latter. This is a direct consequence
of the line flux peaking earlier than the continuum during a flare, as already
discussed (Fig.~\ref{fig:prof}). For example, in a $\sim$day-long observation
(Fourier $f=10^{-5}$--$10^{-4}$), the line may lead the 9 keV continuum 
by as much as 1--3 hours.
The cross-correlation function between the \Ka and 9 keV light curves
is asymmetric, peaking at $\Delta t\approx -10^3$ sec, consistent
with the time lags. The cross-correlation functions plotted, $CCF(\tau)$, were 
normalized
by dividing them by the square root of the product of variances, which means that
for a perfectly correlated signals $CCF(0) = 1$. We can then observe that while two
continuum bands are indeed well correlated, the correlation between 
line and continuum light curves is weaker, resulting in CCF$(0)\approx 0.82$.

The model r.m.s.\ spectra show a minimum at the energy of the \Ka line,
as indeed observed in spectra of Seyfert galaxies (Markowitz et al.\ 2003a).
The minimum is of course a direct consequence of the weakly variable line flux,
which contributes to the count rate in that bin, but only weakly so to the
r.m.s.\ variability. Since in our model the \Ka line is narrow, the reduction of
r.m.s.\ appears only at the line energy bin. Observed lines are broad, therefore
the observed r.m.s.$(E)$ dependence have broad minima around 6--7 keV.

\begin{figure*}
 \parbox{\textwidth}{
  \parbox{0.45\textwidth}{
    \epsfxsize = 0.4\textwidth
    \epsfbox[18 160 620 710]{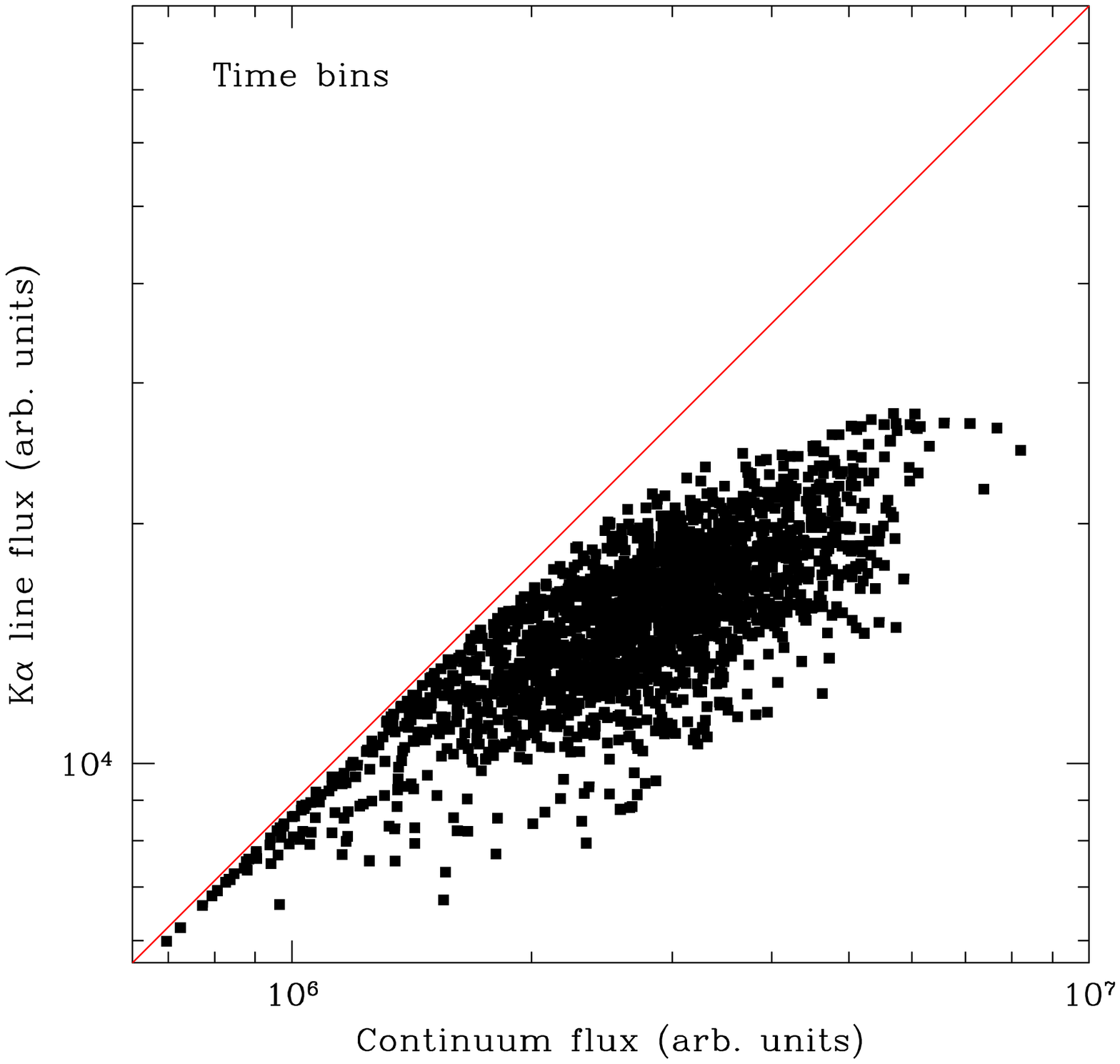}
 }\hfil {
 \parbox{0.45\textwidth}{
    \epsfxsize = 0.4\textwidth
    \epsfbox[18 160 620 710]{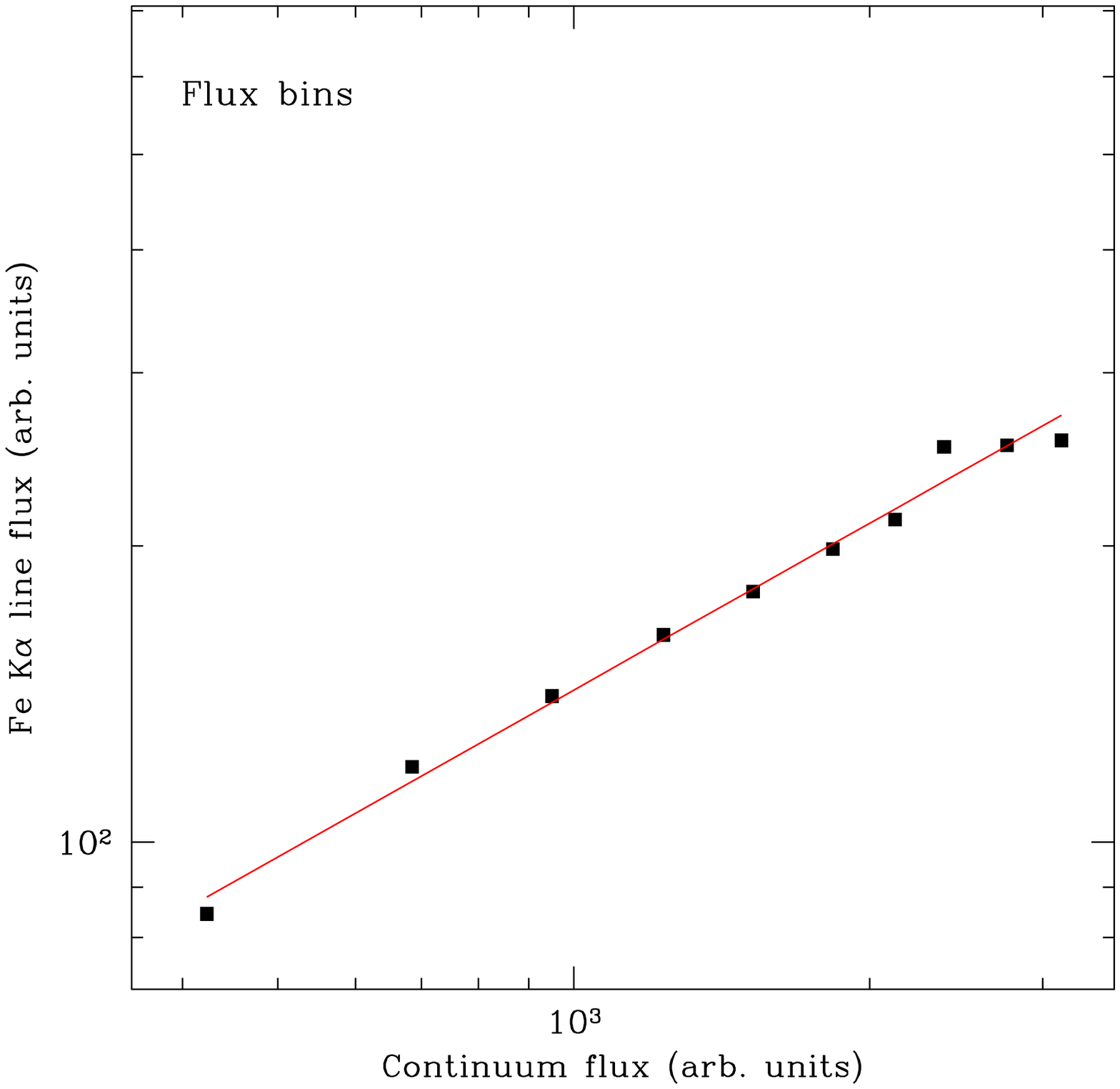}
 }
}}
 \caption{
Line flux vs.\ the 7--30 keV flux. {\it Left panel:\/} each point shows one time bin of 
5000 sec.\ from a $10^7$ sec.\ observation. The line flux follows linearly the continuum flux
for low values of the latter, corresponding to the emission coming from  outer regions
(active region above the cold disc). The higher the continuum flux, the stronger the
deviation from a linear relation, and the bigger the spread of points.
 {\it Right panel:\/} results from spectra binned in
flux. The slope of the best fit line is $\approx 0.56$.
\label{fig:flli}}
\end{figure*}

Fig.~\ref{fig:flli}a shows the zero-lag correlation between the 7--30 keV flux and the
\Ka line flux. Each point represents a spectrum averaged over $\approx 5000$ sec, meant
to roughly represent one RXTE orbit. When the two fluxes are low they are simply linearly
related. This corresponds to emission from the outer regions, $r \ge \rtr$, where
the amplitude of reflection is $\Or=1$. Higher fluxes correspond to emission coming 
at least partially from the
inner region, below $\rtr$, where $\Or<1$, and so the linear relation breaks down.
Fig.~\ref{fig:flli}b shows the $F(7-30)$ vs.\ $F(K\alpha)$ for flux-binned spectra from
the same, $10^7$ sec light curve. In this representations the spatial separation discussed
above is averaged over to some extent, resulting in significantly weaker relation,
with logarithmic slope of $\approx 0.6$.

\section{Discussion}
\label{sec:discuss}

We have computed the variability properties of the Fe \Ka line from X--ray
reprocessing in a propagation model of X--ray emission in accreting compact objects.
The model combines results from various spectral and timing studies of
accreting black holes. The former suggest the geometry of the standard optically thick
accretion disc truncated at a radius larger than the radius of the last stable
orbit (e.g.\ Esin, McClintock \& Narayan 1997; Gierli\'{n}ski et al.\ 1997;
\.{Z}ycki, Done \& Smith 1998; Done \& \.{Z}ycki 1999; 
Zdziarski et al.\ 1999, 2003).
The latter postulate correlated flares (avalanches) and spectral evolution during
flares, in order to explain power spectra and time lags (Poutanen \& Fabian 1999;
Kotov et al.\ 2001).
Combined spectral-timing studies suggest a connection between the geometry and timing
properties (Revnivtsev et al.\ 1999, 2001; \.{Z}ycki 2002, 2003).
Most of the observational results were obtained for black hole binaries, 
but, were possible to conduct, analysis of data for Seyfert galaxies
confirm the general similarity between the two classes of objects 
(e.g.\ spectral studies
by Done et al.\ 2000; Chiang et al.\ 2000; Lubi\'{n}ski \& Zdziarski 2001;
 timing studies of Czerny et al.\ 2001;
Uttley \& M$^{\rm c}$Hardy 2001; Vaughan et al.\ 2003; Markowitz et al.\ 2003b;
Papadakis 2004).

\begin{figure}
 \epsfxsize = 8 cm
 \epsfbox[18 250 620 710]{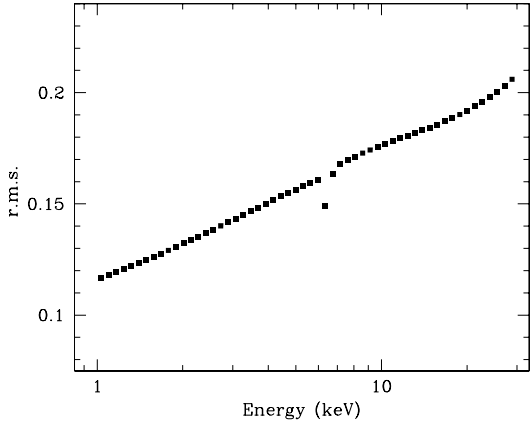}
 \caption{
 Dependence of r.m.s.\ variability on energy. Clear dip at Fe \Ka line energy
 is seen, from the weakly variable line.
\label{fig:rmse}}
\end{figure}

The variability properties of the Fe \Ka line seem to be similar in both classes
of objects in that the line appear to be less variable than the continuum which
drives it, and, where the variability is detected, it does not seem to be clearly
correlated with the continuum. This is contrary to simple(st) ideas, whereby
the continuum and the line are produced in the same region and thus should be
closely related. We note that the observational situation is far from clear,
though. Time resolved spectral analysis would be the most direct method to determine
the variability of the line, but the results may be model dependent
(see detailed discussion in Zdziarski et al.\ 2003). 
Decomposition of counts from a medium energy resolution instrument 
like RXTE/PCA
into the line and continuum depends on the model assumed for both the line and
the reflected continuum. In particular, the effects of relativistic smearing
of the reflected continuum was not taken into account. We note though that
according to Markowitz et al.\ (2003a) the results on line variability in
a sample of Seyfert galaxies were
insensitive to assumptions about the amplitude of the reflected component: 
whether its relative amplitude was fixed, or allowed to vary in accord with
the \Ka line.

Our model does reproduce the reduced variability of the Fe \Ka line, compared to 
the variability of its driving continuum. Line variations may also appear
not exactly correlated with continuum variations, because of the time delay between
the peaks of the line and continuum fluxes.
Both effects are necessary consequences of the adopted geometry, of
a truncated disc with inner hot flow. The same geometry can also explain the hard 
X--ray time lags (Kotov et al.\ 2001; \.{Z}ycki 2003) through spectral evolution
during flares (Poutanen \& Fabian 1999). Quantitatively, our assumed ratio of
heating to cooling rates, $C(r) \propto \lh(r)/\lsoft(r)$, is a much weaker function 
of radius
than could be expected for a compact (size $\ll r$) active region and a sharply
truncated disc. As already mentioned in \.{Z}ycki (2003), in that latter geometry 
the supply of soft photons from the disc would diminish so rapidly that the
predicted energy spectrum would be much too hard to be consistent with the data. 
This implies a {\em gradual\/} disappearance of the cold
disc, which may be an interesting clue as to how the physical process of disc
evaporation proceeds (R\'{o}\.{z}a\'{n}ska \& Czerny 2000).

We emphasize that the reduction of variability of the \Ka line discussed in the present
paper is the same phenomenon as the decreasing reflection amplitude with
Fourier frequency, found in X-ray data of Cyg X-1 and GX 339-4 by Revnivtsev et al.\ (1999, 
2001) and modelled
by \.{Z}ycki (2003). In the present paper this effect was analysed with tools usually
applied to AGN data, which are usually analysed in time domain rather than Fourier domain.
We note, that it does not
seem possible to design the parameters of the model, so that the line flux remains exactly 
constant. This is because, even though the line may be constant during each flare
(assuming $C(r) \propto 1/\lh(r)$), the number of flares active at any time varies,
and this causes variation of the total flux of the line.

A robust feature of the presented model is a connection of the time scale of
the line {\em response\/} to the time scale of continuum variability.
This is because the line flux responds on the time scale related to the
duration of a flare. This in turn has to be chosen such that the observed power
spectra are reproduced. The peaks in $f\times P(f)$ are at rather long time-scales,
much longer than just the light travel time through the region of most efficient 
energy generation. For example, the longest flares in our computations last 
$\approx 10^6$ sec, which in light travel time corresponds to large distance
of $2\approx \times 10^4\,\Rg$ (for $10^7\,\MSun$). This is simply a manifestation
of the well known observational fact that time-scales of maximum X-ray variability 
power are much longer than the naively expected short dynamical time-scale. 
The physical mechanisms of producing the relatively long time scales of variability
are rather unclear, but some interesting possibilities were recently
considered in literature. 3-D magnetohydrodynamical simulations of Narayan, 
Igumenshchev \& Abramowicz (2003) reveal slow ($v\ll v_{\rm ff}$) 
drift of plasma clumps across
lines of magnetic field, which lines are compressed by the accretion flow.
King et al.\ (2004) considered a model involving magnetic dynamos operating locally
in the disc. The longer and larger flares are results of correlations between dynamos
acting in neighboring locations (radii). While individual dynamos operate of short
(dynamical) time scales producing short flares, the correlated behaviour produces
longer lasting, large events. Whatever the exact physical processes are, it is clear
that the accretion flow is highly inhomogeneous and structured, factors that
any realistic modelling should allow for.

In the presented model, the line flux is leading the continuum flux. 
We ignored the light travel time delay of the line
photons after the continuum, but this is unlikely to affect our result, since
the line is supposed to originate close to the location of emission of primary
radiation. The light travel time delay may be estimated as 
$\delta t \sim r\times \Rg/c$ (assuming the height of the emission region $h \sim r$),
which gives $\delta t\sim 10^3$ sec, i.e.\ about the time bin in our simulations.
Obviously, there may be additional effects due to, for example, adjustment of 
properties of the reprocessing medium to the increasing irradiation flux, which 
might affect the result to some extent.

The dependence of r.m.s.\ variability amplitude on energy is a relatively model 
independent demonstration of the reduced variability of the line. Our computations
qualitatively reproduce the minimum of r.m.s.$(E)$ at the energy of the line.
Generally, 
the r.m.s.\ spectra are energy dependent variability amplitude, or, 
equivalently, can be thought of as representing energy spectra of 
the variable component of the spectrum. This can in principle be computed
also in narrow ranges in Fourier frequency (Fourier frequency resolved
spectroscopy, Revnivtsev et al.\ 1999, 2001; \.{Z}ycki 2002, 2003), but 
the quality of AGN data is not sufficient to make use of this technique as yet.

Concluding, the propagation model of high energy emission in the geometry of
a truncated accretion disc provides a framework for understanding many of the
observed spectral and temporal characteristics of X-ray radiation from accreting
black holes.

\section*{Acknowledgments} 
 
This work  was partly supported by grant no.\  2P03D01225
of the Polish State Committee for Scientific Research (KBN).

{}



\begin{thebibliography}{}

 \bibitem[]{}
   Beloborodov A. M., 1999a, ApJ, 510, L123
 \bibitem[]{}
   Beloborodov A. M., 1999b,  in Poutanen J., Svensson R. eds, 
     ASP Conf. Ser. 161, High Energy Processes in Accreting Black Holes, 
     295,  (astro-ph/9901108)
 \bibitem{}
   Chiang J., Reynolds C. S., Blaes O. M., Nowak M. A., Murray N.,
     Madejski G. M., Marshall H. L., Magdziarz P., 2000, ApJ, 528, 292
 \bibitem[]{}
    Czerny B., Lehto H. J., 1997, MNRAS, 285, 365
 \bibitem[]{}
    Czerny B., Niko{\l}ajuk M., Piasecki M., Kuraszkiewicz J., 2001, MNRAS, 325, 865
 \bibitem{}
   Di Salvo T., Done C., \.{Z}ycki P. T., Burderi L., Robba N. R., 2001, ApJ,
    547, 1024
 \bibitem[]{}
   Done C., 2002, Philosophical Transactions of the Royal Society, 360, 1967
  (astro-ph/0203246)
 \bibitem[]{}
   Done C., \.{Z}ycki P. T., 1999, MNRAS, 305, 457
  \bibitem[]{}
   Done C., Madejski G. M., \.{Z}ycki, 2000, ApJ, 536, 213
 \bibitem[]{}
   Esin A. A., McClintock J. E., Narayan R., 1997, ApJ, 489, 865
 \bibitem[]{}
   Fabian A. C., Rees M. J., Stella L., White N. E., 1989, MNRAS, 238, 729
 \bibitem[]{}
   Fabian A. C. et al., 2002, MNRAS, 335, L1
\bibitem []{}
  George I. M., Fabian A. C., 1991, MNRAS, 249, 352
\bibitem
 Gierli\'{n}ski M., Zdziarski A. A., Done C., Johnson W. N., Ebisawa K., Ueda Y., Haardt F.,
Phlips B. F., 1997, MNRAS, 288, 958
\bibitem []{}
  Gilfanov M., Churazov E.,  Revnivtsev M., 2000, Proc.\  5th 
   Sino-German Workshop on Astrophysics, SGSC Conference Ser., Vol. 1,
  China Science \& Technology Press, Beijing, p. 114 (astro-ph/0002415)
 \bibitem[]{}
   Inoue H., Matsumoto C., 2001, AdSpRes, 28, 445
 \bibitem[]{}
  King A. R., Pringle J. E., West R. G., Livio M., 2004, MNRAS, 348, 111
 \bibitem[]{}
   Kotov O., Churazov E., Gilfanov M., 2001, MNRAS, 327, 799
 \bibitem[]{}
   Lightman A. P., White T. R., 1988, ApJ, 335, L57
 \bibitem[]{}
   Lubi\'{n}ski P., Zdziarski A. A., 2001, MNRAS, 323, L37
\bibitem[]{}
    Markowitz A., Edelson R., Vaughan S., 2003a, ApJ, 598, 935
\bibitem[]{}
    Markowitz A. et al., 2003b, ApJ, 593, 96
  \bibitem[]{}
    Martocchia A., Matt G., Karas V., Belloni T. Feroci M., 2002, A\&A, 387, 215
  \bibitem[]{}
     Matt G., Perola G., Piro L., 1991, A\&A, 247, 25
  \bibitem[]{}
     McClintock J. E., Remillard R. A., 2003, preprint (astro-ph/0306213)
 \bibitem[]{}
    Misra R., 2000, ApJ, 529, L95
 \bibitem[]{}
   Miyamoto S.,  Kitamoto S., Mitsuda K., Dotani T., 1988, Nat, 336, 450
\bibitem[]{}
   Narayan R., Igumenshchev I. V., Abramowicz M. A., 2003, PASJ, 55, L69
  \bibitem[]{}
   Nayakshin S., Kazanas D., Kallman T. R., 2000, ApJ, 537, 833
 \bibitem[]{}
   Nowak M. A., Wilms J., Vaughan B. A., Dove J. B, Begelman M. C., 1999, 
             ApJ, 515, 726
 \bibitem[]{}
  Papadakis Y., 2004, MNRAS, 348, 207
 \bibitem[]{}
   Poutanen J., 1999, in Abramowicz M. A., Bj\"{o}rnsson G., Pringle J. E., 
   eds, Theory of Black Hole Accretion Discs. CUP, Cambridge, p. 100
   (astro-ph/9805025)
 \bibitem[]{}
   Poutanen J., 2001, AdSpR, 28, 267 (astro-ph/0102325)
 \bibitem[]{}
   Poutanen J., Fabian A. C., 1999, MNRAS, 306, L31
\bibitem[]{}
   Poutanen J., Krolik J. H., Ryde F., 1997, MNRAS, 292, L21
 \bibitem[]{}
   Revnivtsev M., Gilfanov M., Churazov E., 1999, A\&A, 347, L23
 \bibitem[]{}
   Revnivtsev M., Gilfanov M., Churazov E., 2001, A\&A, 380, 520
 \bibitem[]{}
   Reynolds C. S., 2000, ApJ, 533, 811
 \bibitem[]{}
   Reynolds C. S., Nowak M. A., 2003, Physics Reports, 377, 389
 \bibitem[]{}
   R\'{o}\.{z}a\'{n}ska A., Czerny B., 2000, A\&A, 360, 1170
 \bibitem[]{}
   Shakura N. I., Sunyaev R. A., 1973, A\&A, 24, 337
 \bibitem[]{}
   Tanaka Y. et al., 1995, Nat, 375, 659
 \bibitem[]{}
   Uttley P., 2004, MNRAS, 347, L61
 \bibitem[]{}
   Uttley P., M$^{\rm c}$Hardy I. M., 2001, MNRAS, 323, L26
 \bibitem[]{}
  van der Klis M., 1995, in Lewin W. H. G., van Paradijs J., 
      van den Heuvel E. P. J., eds, X--ray binaries, Cambridge Univ. Pres,
       Cambridge, p.\ 252
 \bibitem[]{}
   Vaughan B. A., Nowak M. A., 1997, ApJ, 474, 43
 \bibitem[]{}
  Vaughan S., Edelson R., 2001, ApJ, 548, 694
 \bibitem[]{}
   Zdziarski A. A., Johnson W. N., Magdziarz P., 1996 ApJ, 283, 193
 \bibitem[]{}
   Zdziarski A. A., Lubi\'{n}ski P., Smith D. A., 1999, MNRAS, 303, L11
 \bibitem[]{}
   Zdziarski A. A., Lubi\'{n}ski P., Gilfanov M., Revnivtsev M., MNRAS, 342, 355
 \bibitem[]{}
    \.{Z}ycki P. T., 2002, MNRAS, 333, 800 
 \bibitem[]{}
    \.{Z}ycki P. T., 2003, MNRAS, 340, 639
 \bibitem[]{}
    \.{Z}ycki P. T., Czerny B., 1994, MNRAS, 266, 653
 \bibitem[]{}
    \.{Z}ycki P. T., R\'{o}\.{z}a\'{n}ska A., 2001, MNRAS, 325, 197
 \bibitem[]{}
    \.{Z}ycki P. T., Done C.,  Smith D. A., 1998, ApJ, 496, L25 

\label{lastpage}

\end{thebibliography}
\end{document}